\documentclass[10pt,conference]{IEEEtran}
\IEEEoverridecommandlockouts
\usepackage{cite}
\usepackage{amsmath,amssymb,amsfonts}
\usepackage{algorithmic}
\usepackage{graphicx}
\usepackage{textcomp}
\usepackage{url}
\usepackage{multirow}
\usepackage{tcolorbox}
\usepackage{booktabs}
\usepackage{tcolorbox}
\tcbuselibrary{skins}
\usepackage{multirow}
\usepackage{bbding}
\usepackage{utfsym}
\usepackage{ulem}
\usepackage{balance}
\usepackage{wasysym}
\usepackage{hyperref}
\usepackage{multirow}
\usepackage{makecell}
\usepackage{pifont}
\usepackage{subfigure}
\usepackage{amsmath}
\usepackage{algorithm}
\usepackage{algorithmic}

\usepackage[utf8]{inputenc}
\usepackage[T1]{fontenc}

\def\eg{\textit{e.g.,} }
\def\ie{\textit{i.e.,} }

\newtcolorbox{AcademicBox}[1][]{academicbox=#1}
\definecolor{SoftBlue}{RGB}{135, 206, 250}  
\definecolor{SoftOrange}{RGB}{255, 224, 178} 
\definecolor{SoftGreen}{RGB}{144, 238, 144}  
\definecolor{CorrectGreen}{RGB}{76, 175, 80} 
\definecolor{ErrorRed}{RGB}{211, 47, 47} 

\def\BibTeX{{\rm B\kern-.05em{\sc i\kern-.025em b}\kern-.08em
    T\kern-.1667em\lower.7ex\hbox{E}\kern-.125emX}}

\tcbset{
  my box/.style={
    enhanced,
    colframe=#1!80,
    colback=#1!10,
    attach boxed title to top left={xshift=0.2cm, yshift=-0.2cm},
    boxed title style={
      colback=#1!80,
      outer arc=0pt,
      arc=0pt,
      top=0pt,
      bottom=0pt,
    },
  },
}

\newtcolorbox{resultRQ}[1]{
  my box=black,
  title=#1,
  boxrule=1.2pt,top=6pt,bottom=3.5pt,left=6pt,right=6pt
}

\begin{document}

\title{A First Look at Package-to-Group Mechanism: An Empirical Study of the Linux Distributions}



\author{\IEEEauthorblockN{Dongming Jin}
\IEEEauthorblockA{
Key Lab of High Confidence Software \\
Technology, MoE (Peking University) \\
Beijing, China \\
dmjin@stu.pku.edu.cn}
\and
\IEEEauthorblockN{Nianyu Li, Kai Yang}
\IEEEauthorblockA{
ZGC National Laboratory \\
Beijing, China \\
li\_nianyu@pku.edu.cn\\
yangkai@mail.zgclab.edu.cn}
\and
\IEEEauthorblockN{Minghui Zhou, Zhi Jin* \thanks{*Corresponding authors}}
\IEEEauthorblockA{
Key Lab of High Confidence Software \\
Technology, MoE (Peking University) \\
Beijing, China \\
mhzhou@pku.edu.cn, zhijin@pku.edu.cn}
}

\maketitle

\begin{abstract}
Reusing third-party software packages is a common practice in software development. As the scale and complexity of open-source software (OSS) projects continue to grow (e.g., Linux distributions), the number of reused third-party packages has significantly increased. Therefore, maintaining effective package management is critical for developing and evolving OSS projects. To achieve this, a package-to-group mechanism (P2G) is employed to enable unified installation, uninstallation, and updates of multiple packages at once. 
To better understand this mechanism, this paper takes Linux distributions as a case study and presents an empirical study focusing on its application trends, evolution patterns, group quality, and group tendency. By analyzing 11,746 groups and 193,548 packages from 89 versions of 5 popular Linux distributions and conducting questionnaire surveys with Linux practitioners and researchers, we derive several key insights. Our findings show that P2G is increasingly being adopted, particularly in popular Linux distributions. P2G follows six evolutionary patterns (\eg splitting and merging groups). Interestingly, packages no longer managed through P2G are more likely to remain in Linux distributions rather than being directly removed. Besides, we propose a metric called {\sc GValue} to evaluate the quality of groups and identify issues such as inadequate group descriptions and insufficient group sizes.  We also summarize five types of packages that tend to adopt P2G, including graphical desktops and networks. To the best of our knowledge, this is the first study focusing on the P2G mechanisms. We expect our study can assist in the efficient management of packages and reduce the burden on practitioners in rapidly growing Linux distributions and other open-source software projects. 
\end{abstract}


\section{Introduction}
The development of modern software systems heavily relies on code reuse, often achieved through the use of third-party packages~\cite {Lim94}. Utilizing packages to build software can reduce development costs and enhance productivity~\cite{MohagheghiCKS04},~\cite{MohagheghiC07}. Due to these advantages, third-party packages are also widely used in open-source software (OSS) projects~\cite{Wang0HSX0WL20}. However, with the rapid rise in popularity of OSS projects due to their low cost and high quality~\cite{FrakesK05},~\cite{west2006challenges}, their scale and complexity have also significantly increased~\cite{torvalds2001just},~\cite{mockus2000case}, resulting in a sharp rise in the reuse of third-party packages. For instance, CentOS version 7 contains 14,479 packages, while version 3.7 released in 2006 has only 1,275 packages~\cite{web:centos}. This growth presents a critical challenge: effectively managing packages to ensure the ongoing development, continuous evolution, and long-term success of OSS projects.

Managing and reusing packages in OSS projects is particularly challenging~\cite{IshioKKGI16}, particularly for those requiring long-term maintenance and widespread adoption~\cite{ButlerGLBMGFL20}. One of the difficulties arises from the fact that OSS projects often undergo frequent feature changes, and each feature typically relies on multiple packages for support.  As a result, when a feature is modified, maintainers must handle numerous related packages. For instance, removing \textit{Tomcat} support in a Linux distribution requires uninstalling packages such as \textit{tomcat-admin-webapps}, \textit{tomcat-native}, and \textit{tomcat-webapps}. Manually identifying and removing these packages individually increases the workload and pressure on OSS developers and maintainers. 

\begin{figure}
    \centering
    \includegraphics[width=\linewidth]{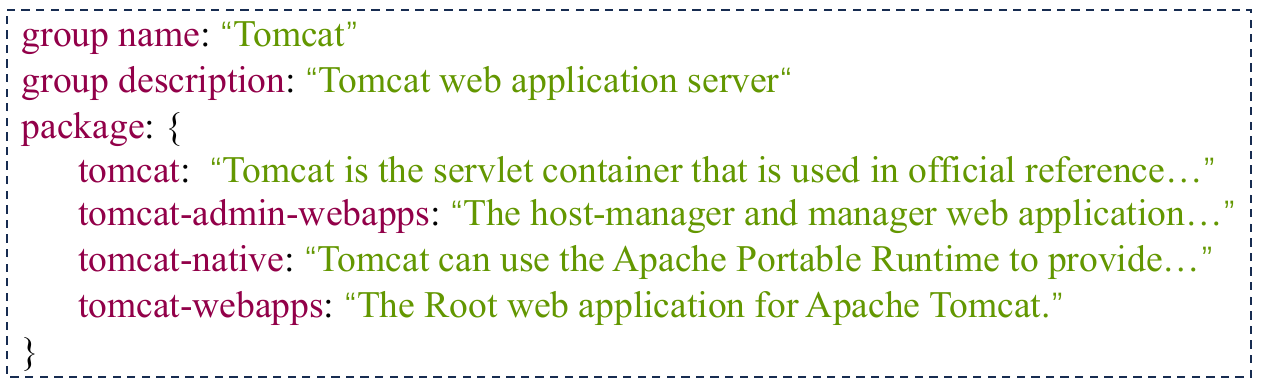}
    \caption{An example of the P2G mechanism in CentOS 7}
    \label{fig:teaser}
\end{figure}

To tackle this issue, many OSS projects employ package managers (\eg RPM~\cite{ewing1996rpm} and APT~\cite{web:apt}) to bundle multiple packages that support a specific feature into a single unit (\ie a group), as shown in Figure \ref{fig:teaser}. Installing or uninstalling a group can automatically add or remove all packages in this group. In this paper, we refer to such activities as a Package-to-Group (P2G) mechanism. Although this mechanism has been adopted in many OSS projects for years, developers and maintainers still encounter numerous issues. For instance, the \textit{networkmanager} package is placed in the \textit{GNOME} group, which creates inconvenience for maintainers during uninstallation~\cite{web:P2Gissue}, leading to unnecessary dependencies being removed. Therefore, investigating the P2G mechanism can better support OSS project developers and maintainers in practicing this mechanism and managing packages more efficiently.

In our literature review, we found that there has been no prior research on the P2G mechanism. Previous studies related to package management mainly focus on dependency analysis~\cite{DecanMZR22},~\cite{JafariCSA23}, freshness evaluation~\cite{LegayDM20},~\cite{LegayDM21}, and malicious detection~\cite{HalderBMJI0IAR024},~\cite{SejfiaS22}. Additionally, some studies have discussed package layering~\cite{tan2023case} and classification~\cite{jing2023classifying}. Their core concepts are similar to P2G, as they all involve dividing packages into different sets (\ie groups, layers, or categories). Compared to P2G, these approaches employ broader, more coarse-grained divisions. This leads to large groups containing an excessive number of packages, which causes inconveniences in managing or updating packages when a particular feature of the OSS project changes.

In this paper, we take Linux distributions as a case study to investigate the P2G mechanism in an evolution context. Linux distributions are chosen due to their representativeness for OSS projects and their widespread adoption and vast ecosystem of packages~\cite{TanZ020}. Specifically, we start by understanding the application trends of this mechanism and analyzing the evolutionary patterns across \textbf{previous versions}. Then we propose a metric called {\sc GValue} to evaluate the quality and identify issues in \textbf{current versions}. Finally, we explore the group tendency to support P2G in \textbf{future versions}. In particular, this paper addresses the following research questions (RQs):

\textbf{RQ1: How common is the use of the P2G mechanism?} We carefully select 5 Linux distributions and identified groups, packages within those groups, and the total number of packages in each distribution. We analyze the trend of adopting this mechanism over versions, the distributions of packages within P2G, and the correlation between P2G adoption and distribution popularity. Our analysis shows that while P2G is becoming increasingly common, especially in popular distributions, it still accounts for only a small proportion of total packages across Linux distributions.

\textbf{RQ2: What are the evolutionary patterns in the P2G mechanism?} We address this from two perspectives: a content-based summary of change patterns and a quantitative analysis of package flows. Specifically, we manually analyze the addition and removal of groups across 89 consecutive versions of five Linux distributions to summarize the P2G change patterns and calculate the number of affected packages in each flow scenario. Finally, we build a taxonomy of these change patterns and find that packages no longer adopting P2G are more likely to remain in Linux distributions.

\textbf{RQ3: What is the quality of the current groups in the P2G mechanism?} To evaluate group quality, we establish a multi-dimensional feature method called {\sc GValue}. To validate its effectiveness, we randomly select 5 groups to design a questionnaire survey for manual scoring, followed by a correlation analysis~\cite{de2016comparing} with {\sc GValue}. Our results show that 16\% of the groups exhibit poor quality (\ie {GValue} score below 0.2) across Linux distributions. Additionally, we identify several existing issues on P2G adoption from an analysis of 413 groups in the latest version. 

\textbf{RQ4: What types of packages do developers tend to adopt the P2G mechanism?} We conduct a topic analysis on group description and identify 5 categories (\eg Graphical Desktop) that are more likely to adopt P2G. Furthermore, we extract key information from package descriptions and compare them between packages within and without P2G. This analysis provides insights for developers and maintainers in deciding whether a package should adopt P2G.

We summarize our contributions in this paper as follows.
\begin{itemize}
    \item We provide a comprehensive analysis of the P2G mechanism covering its application trends, evolutionary patterns, group quality, and group tendency.
    \item We summarize various evolutionary patterns in groups, offering insights that guide OSS project maintainers in adopting P2G in future updates.
    \item We propose a method for measuring the quality of groups and identifying current issues, which can enhance quality control for the adoption of P2G.
    \item We discover 5 categories of packages where developers frequently adopt P2G, offering valuable guidance for Linux maintainers in deciding whether to adopt P2G for a specific package.
\end{itemize}

\textbf{Data Availability}. We have open-sourced our replication package~\cite{web:code}, including the dataset and source code, to facilitate other researchers and practitioners in replicating our work and validating their research.  
\section{Background and related work} 
In this section, we first review related works on package management in Linux distributions. Then, we provide a brief introduction to the context of the P2G mechanism. 

\subsection{Related Work}
As OSS projects evolve, they are becoming larger and more complex~\cite{lehman1997metrics}, leading to a significant rise in the reuse of third-party packages. Developers and maintainers of OSS projects face increasing pressure to manage these packages, which can strain their capacity~\cite{AlfadelCS21}. 

Various theories and strategies have been proposed to reduce the workload of package management and enhance the productivity of OSS projects~\cite{MuhammadRH19}. Wang et al.~\cite{wang2015graph} introduced a graphical method to establish dependency relations within a Linux distribution and analyzed the complex graph of related attributes. Sousa et al.~\cite{de2009analysis} divided packages into communities based on their dependencies, recommending that the same team maintain packages in the same community according to the strength of their relationships. Legay et al.~\cite{LegayDM21} evaluated and compared the freshness of 890 packages across six mainstream Linux distributions. Halder et al.~\cite{HalderBMJI0IAR024} proposed a metadata-based model for malicious package detection. Thus, previous studies mainly focus on package dependency analysis, freshness evaluation, and malicious detection. There is a lack of in-depth research on the package-to-group mechanism in OSS projects.

Although no prior studies directly focus on the P2G mechanism, some related works exist ~\cite{tan2023case}~\cite{jing2023classifying}. They all involve classifying multiple packages into distinct sets. Tan et al. ~\cite{tan2023case} suggested that a clear package hierarchy can facilitate the rapid construction and maintenance of Linux distributions. They proposed an efficient and accurate package layering algorithm that divides packages into four layers. However, their approach categorizes packages into only four levels, resulting in an excessive number of packages within the same level. While this approach aids in building integrated Linux and retrieving specific packages, its overall effectiveness is limited. Jing et al.~\cite{jing2023classifying} argue that package categorization can guide Linux developers in selecting components and provide guidance during the construction process. They created a categorized dataset of packages and outlined selection boundaries for component architecture, offering useful insights for the construction process. However, this work only focuses on a single Linux distribution, and the number of categories remains too few to handle the complexity of OSS projects.

To further advance these works, this paper focuses on the P2G mechanism in Linux distributions, investigating its application trends, evolutionary patterns, group quality, and group tendency. This work aims to assist developers and maintainers in the OSS projects efficiently manage packages by adopting the P2G mechanism.

\subsection{Package-to-group Mechanism}
The package-to-group mechanism bundles multiple packages with similar purposes into a group, as shown in Figure \ref{fig:teaser}. This mechanism primarily simplifies the installation and management, enhancing efficiency for developers and maintainers. Different Linux distributions implement this mechanism in different ways. For instance, Debian and Debian-based Linux distributions (\eg Ubuntu) use \textit{metapackages}~\cite{web:metapackage}, which defines groups of multiple packages based on their purposes. This allows developers and maintainers to integrate a specific functional feature in one step, \ie install a corresponding group with all relevant packages. In Red Hat and its derivatives (\eg Fedora and CentOS), this mechanism is based on the concept of \textit{groups}. These groups can be defined by a configuration file, which lists all groups and their associated packages. Maintainers can easily install or uninstall groups, making it easier to manage large sets of packages.

The P2G mechanism not only helps maintainers more easily locate and install the packages they need but also facilitates the maintenance of OSS projects. Additionally, this mechanism can simplify security updates and upgrades, as OSS projects can be targeted at specific groups rather than updating each package individually. This is particularly important in managing servers and large enterprise environments.

\section{Overview}
In this section, we present the design of our empirical study and provide an overview of the selected Linux distributions.

\subsection{Study Design}
\begin{figure}
    \centering
    \begin{minipage}{0.9\linewidth}
		\centerline{\includegraphics[width=\textwidth]{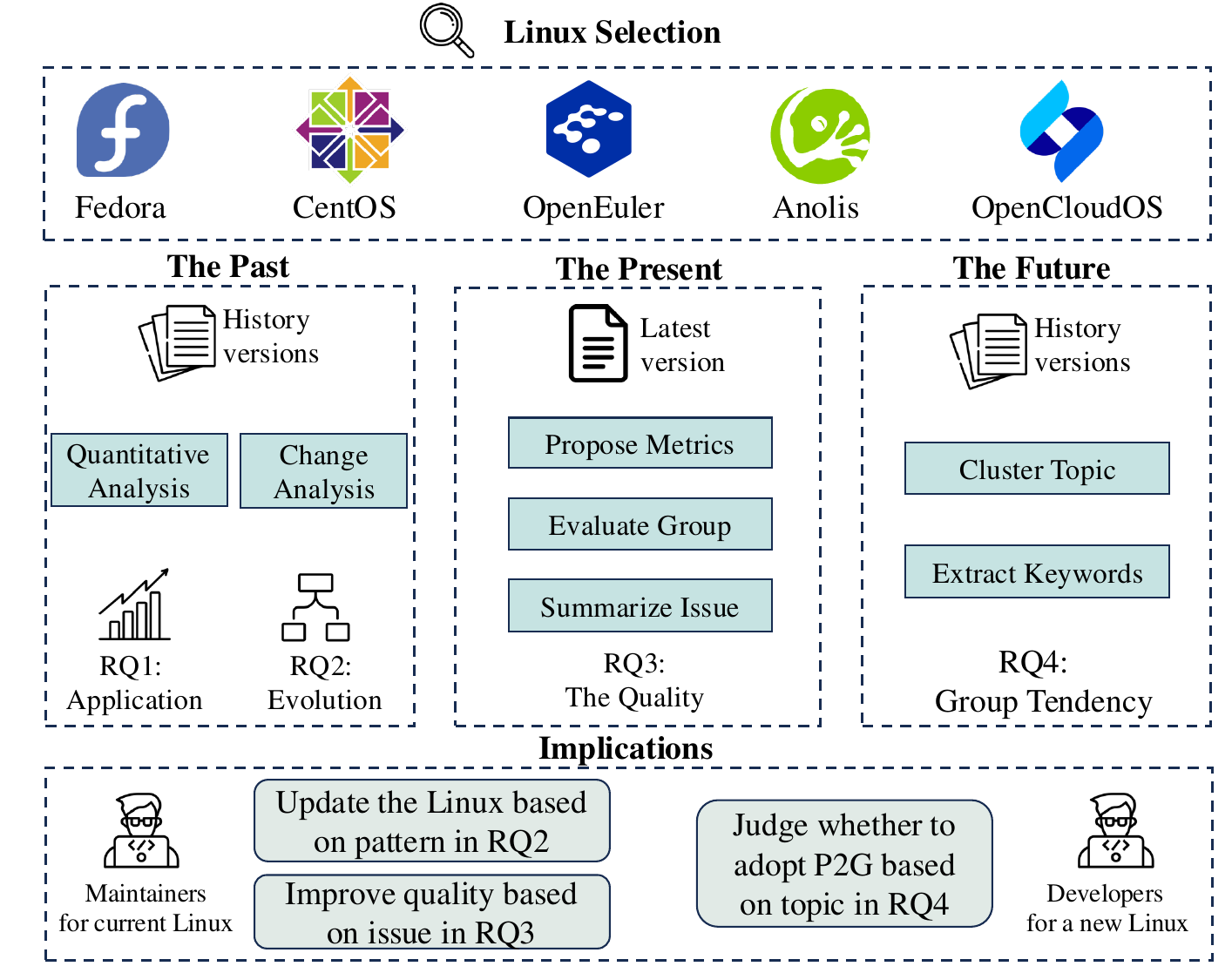}}
    \end{minipage}
    \caption{Overview of our study}
    \label{fig:study-design}
\end{figure}

Figure \ref{fig:study-design} provides an overview of our empirical study, which focuses on five Linux distributions (Section \ref{sec:linux}). Our study follows the evolutionary process and can be divided into three phases (\ie the past, the present, and the future). In the past phase, we conduct quantitative analysis and change analysis on historical versions to investigate the P2G's application (\hyperref[sec:rq1]{RQ1}) and evolution (\hyperref[sec:rq2]{RQ2}). In the present phase, we assess the latest version's quality (\hyperref[sec:rq3]{RQ3}) by proposing metrics, evaluating groups, and summarizing issues. In the future phase, we conduct a topic analysis for group descriptions and key information extraction from package descriptions to explore group tendency (\hyperref[sec:rq4]{RQ4}). The practical implications of our study include providing current Linux maintainers with update recommendations based on evolutionary patterns and quality issues and helping Linux developers determine whether to adopt P2G based on group tendency.

\begin{table}[]
    \centering
     \caption{Studied Linux distributions in our work}
    \begin{tabular}{llll}
\hline
\textbf{Linux}       & \textbf{version(\#)}            & \textbf{company}            & \textbf{mirror}\\ \hline
Fedora      & 7$\sim$39  (33)        & Red Hat            & \href{https://archives.fedoraproject.org/pub/fedora/linux/}{Fedora Project} \\
CentOS      & 3.7$\sim$8.0 (34)           & Red Hat            & \href{https://archives.fedoraproject.org/pub/fedora/linux/}{CentOS Mirror}\\
OpenEuler   & 20.03$\sim$23.09 (7)          & Hua Wei            & \href{https://mirrors.tuna.tsinghua.edu.cn}{Tsinghua Mirror}\\
Anolis      & 7.7$\sim$8.9 (8)             & OpenAnolis         & \href{https://mirrors.aliyun.com}{Aliyun Mirror}\\
openCloudOS & 7$\sim$9.0 (7)               & China Electronics  & \href{https://mirrors.opencloudos.tech}{CloudOS Mirror}\\ \hline
\end{tabular}
    \label{tab:linux_os}
\end{table}

\subsection{Studied Linux Distributions} \label{sec:linux}
There are many package management tools for Linux, such as RPM for Red Had-based distributions and DEB for Debian-based distributions. Since RPM has become one of the recognized standards in Linux distributions~\cite{ewing1996rpm}, we rank RPM-based distributions according to developers' interests~\cite{web:popularity} and select the two most popular open-source distributions, \ie Fedora and CentOS. They are widely used general-purpose distributions. To provide broader coverage of Linux across different scenarios, we include another three distributions tailored for specific cases, \ie OpenEuler~\cite{zhou2022openeuler}, Anolis~\cite{losos2009anolis} and OpenCloudOS~\cite{web:opencloudos}. OpenEuler is optimized for cloud and edge computing and is suitable for industry-specific applications. Anolis focus on enterprise-level applications and is a popular choice for data centers. OpenCloudOS is designed for could computing environments and is suitable for cloud service providers. Additionally, they are all representative distributions from different companies. Table \ref{tab:linux_os} shows the studied Linux distributions and
their versions in our study. We obtained package and group information for each distribution from official or widely available mirrors.

\section{RQ1:Application Trends} \label{sec:rq1}
We aim to examine the evolving trends in the application of the P2G mechanism across different versions of Linux distributions through quantitative analysis.

\subsection{Methodology}
We answer this RQ from three aspects: 1) the trend in the number of groups, the number of packages adopting P2G, and the total number of packages; 2) the distribution of packages adopting P2G within a Linux distribution; 3) the relationship between the adoption of P2G and distribution popularity. The analysis is based on the selected Linux distributions and release versions in Section \ref{sec:linux}. To decide whether a package adopts P2G in a specific release version, we consider whether it is included in a group in that version.

\subsection{Results}
\begin{figure}[htb]
	\centering
	\begin{minipage}{0.88\linewidth}
		\centerline{\includegraphics[width=\textwidth]{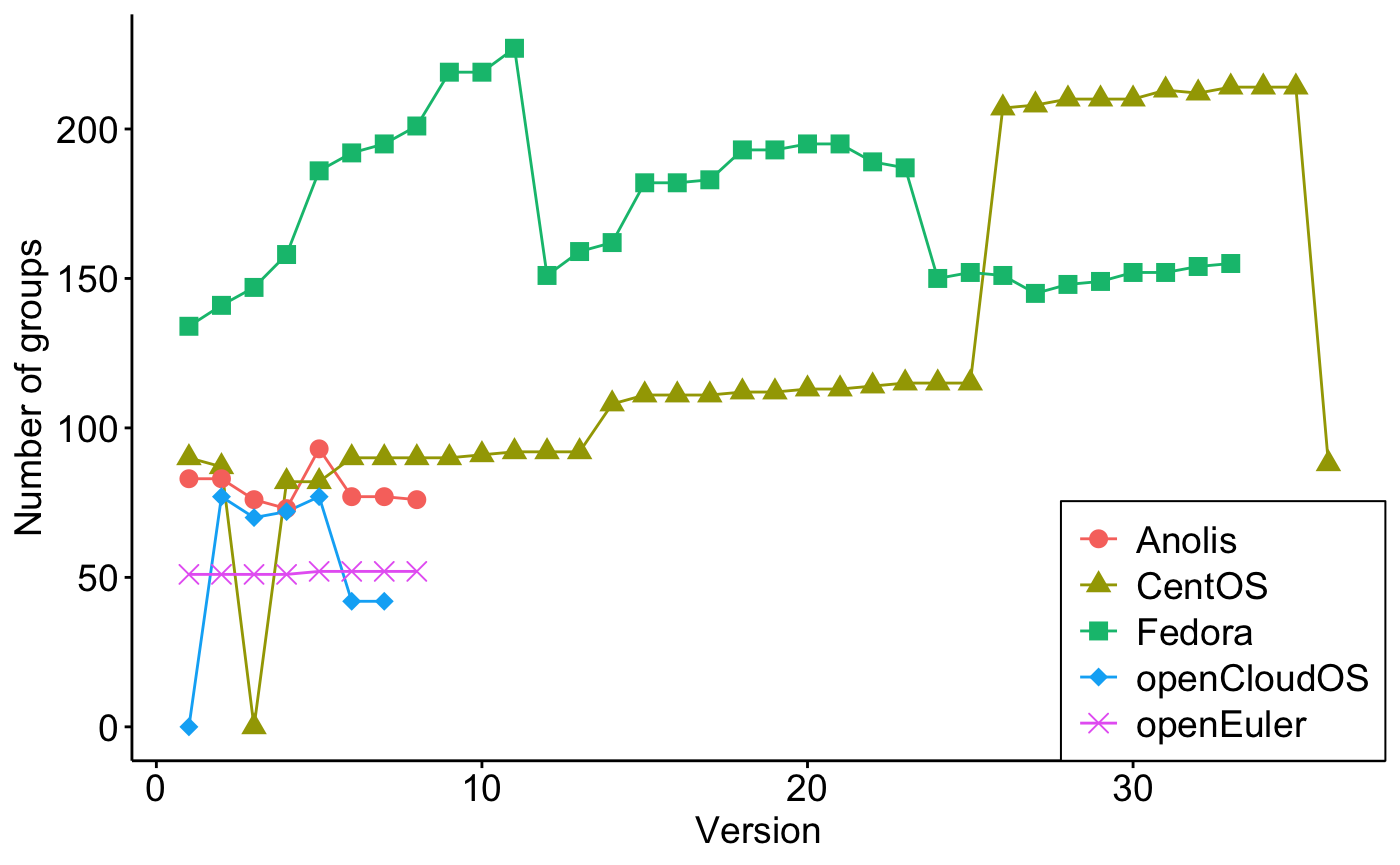}}
		\centerline{Trend in number of groups}
	\end{minipage}
	\vspace{10pt} 
	\begin{minipage}{0.88\linewidth}
		\centerline{\includegraphics[width=\textwidth]{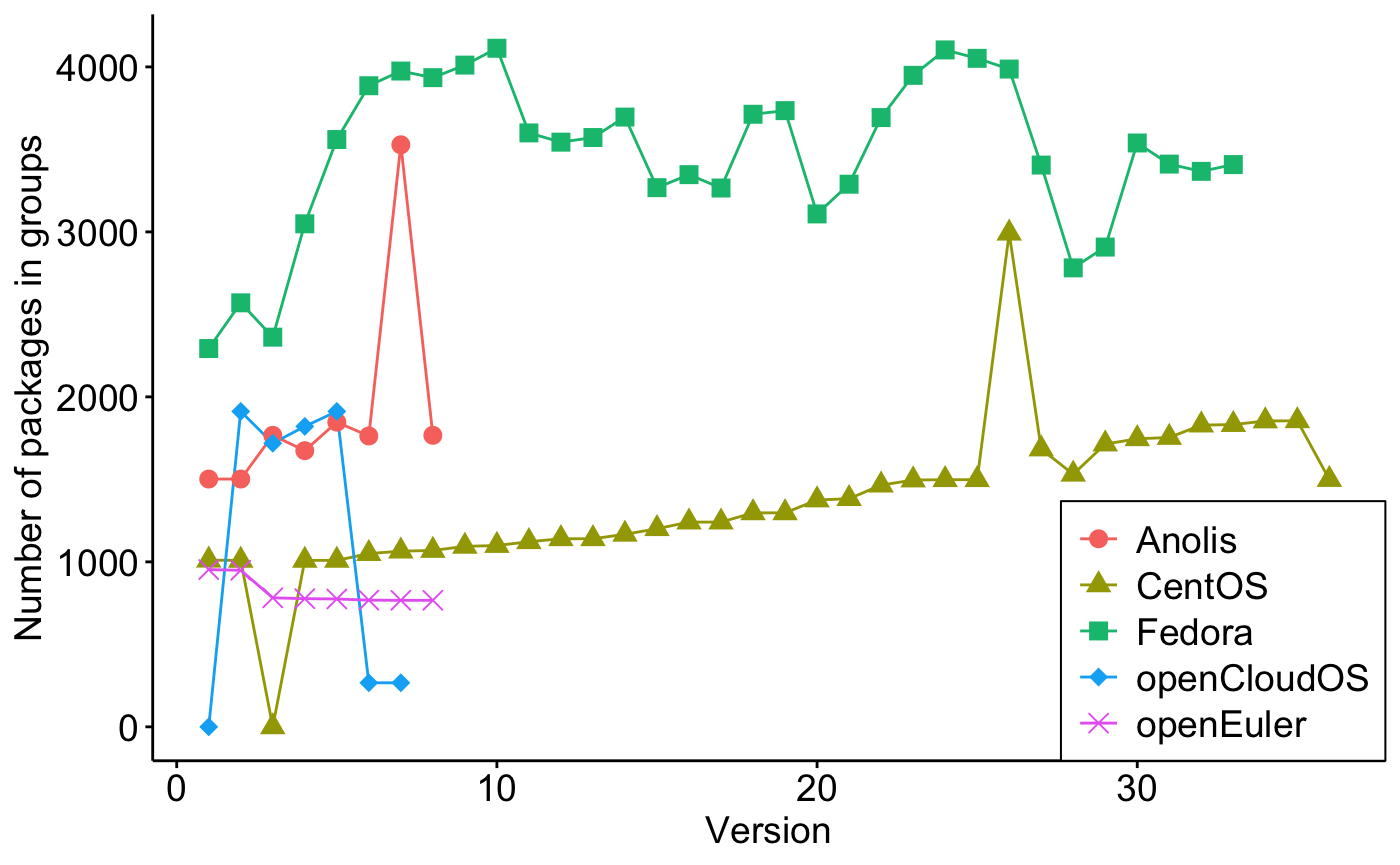}}
		\centerline{Trend in number of packages applied P2G}
	\end{minipage}
	\vspace{10pt} 
	\begin{minipage}{0.88\linewidth}
		\centerline{\includegraphics[width=\textwidth]{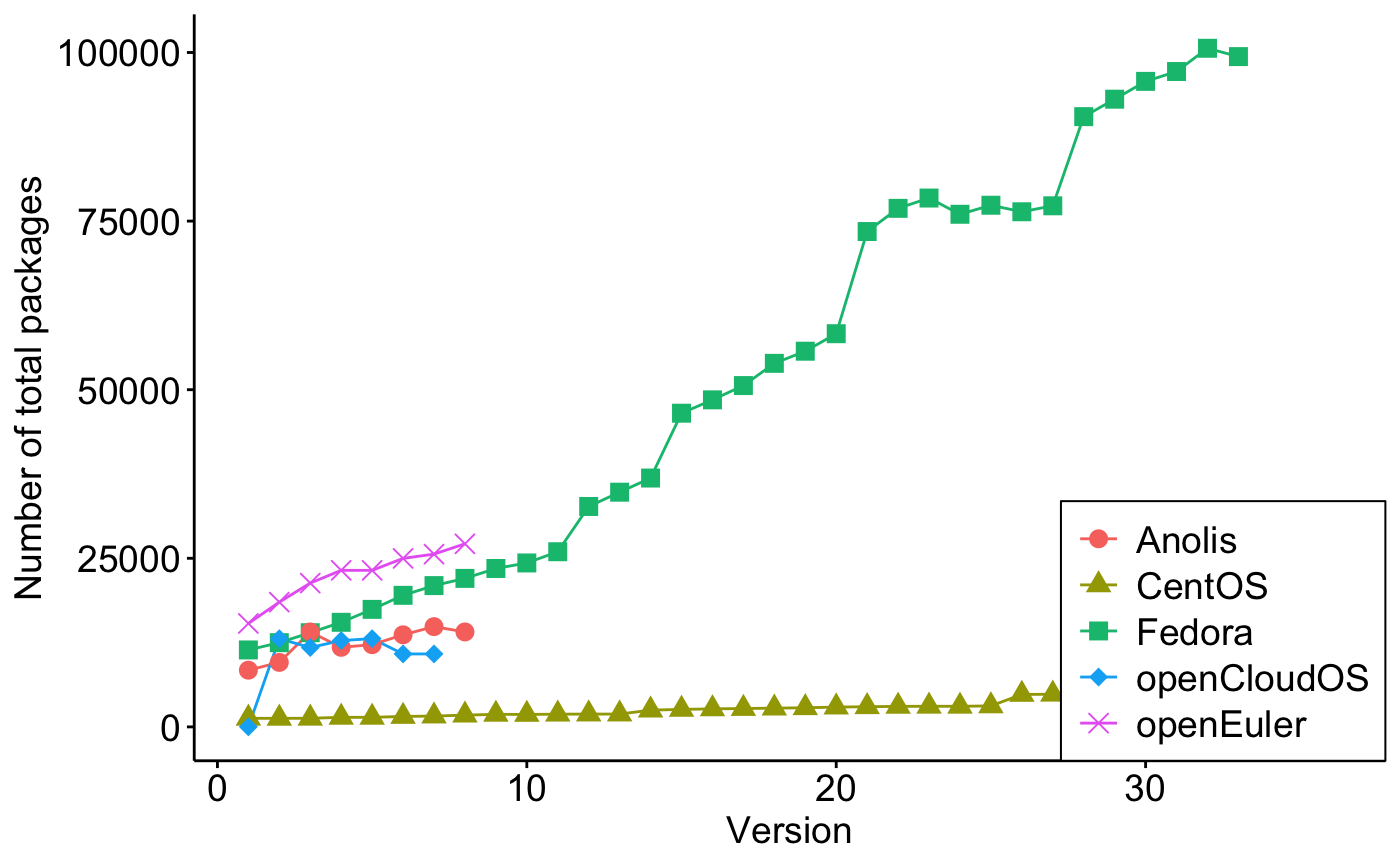}}
		\centerline{Trend in number of packages}
	\end{minipage}
	\vspace{-0.2cm}
	\caption{Trend of the application of the mechanism of P2G}
	\label{rq1-trend}
\end{figure}

\textbf{Application Trends.} Figure \ref{rq1-trend} shows the number of groups, packages adopting P2G, and total packages across all historical release versions. \ding{182} We observe that the number of groups remains relatively stable whether in a complex Linux distribution (\eg Fedora) or a relatively simplified distribution (\eg OpenEuler). This suggests that Linux distributions have a relatively fixed group structure. \ding{183} However, most Linux distributions show an increase in the number of packages adopting P2G, especially in more innovative and developer-oriented distributions (\eg Fedora). In contrast, in highly stable scenarios (\eg enterprise or server environments), Linux distributions (\eg openEuler) show less variation, as they prioritize reliability and compatibility over frequent changes. \ding{184} The number of total packages continues to increase with version updates. This reflects that Linux distributions continuously offer additional features and support in response to enhanced functionality and growing user requirements. For example, the number of packages in Fedora increased nearly tenfold, from 11,410 to 103,176.

\begin{figure}[htb]
	\centering
	\begin{minipage}{0.85\linewidth}
		\centerline{\includegraphics[width=\textwidth]{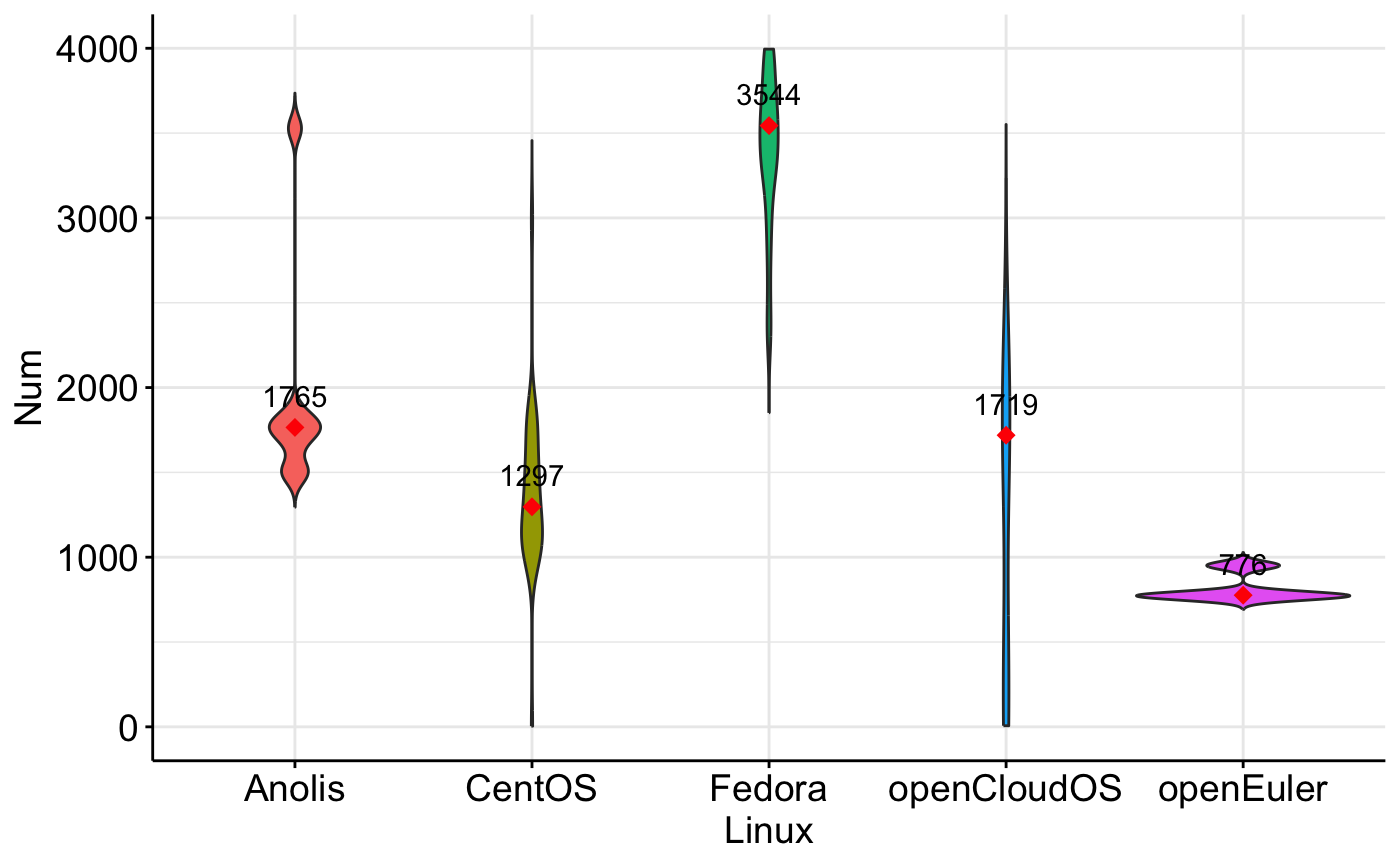}}
		\centerline{Distribution of number per version}
	\end{minipage}
	\vspace{10pt} 
	\begin{minipage}{0.85\linewidth}
		\centerline{\includegraphics[width=\textwidth]{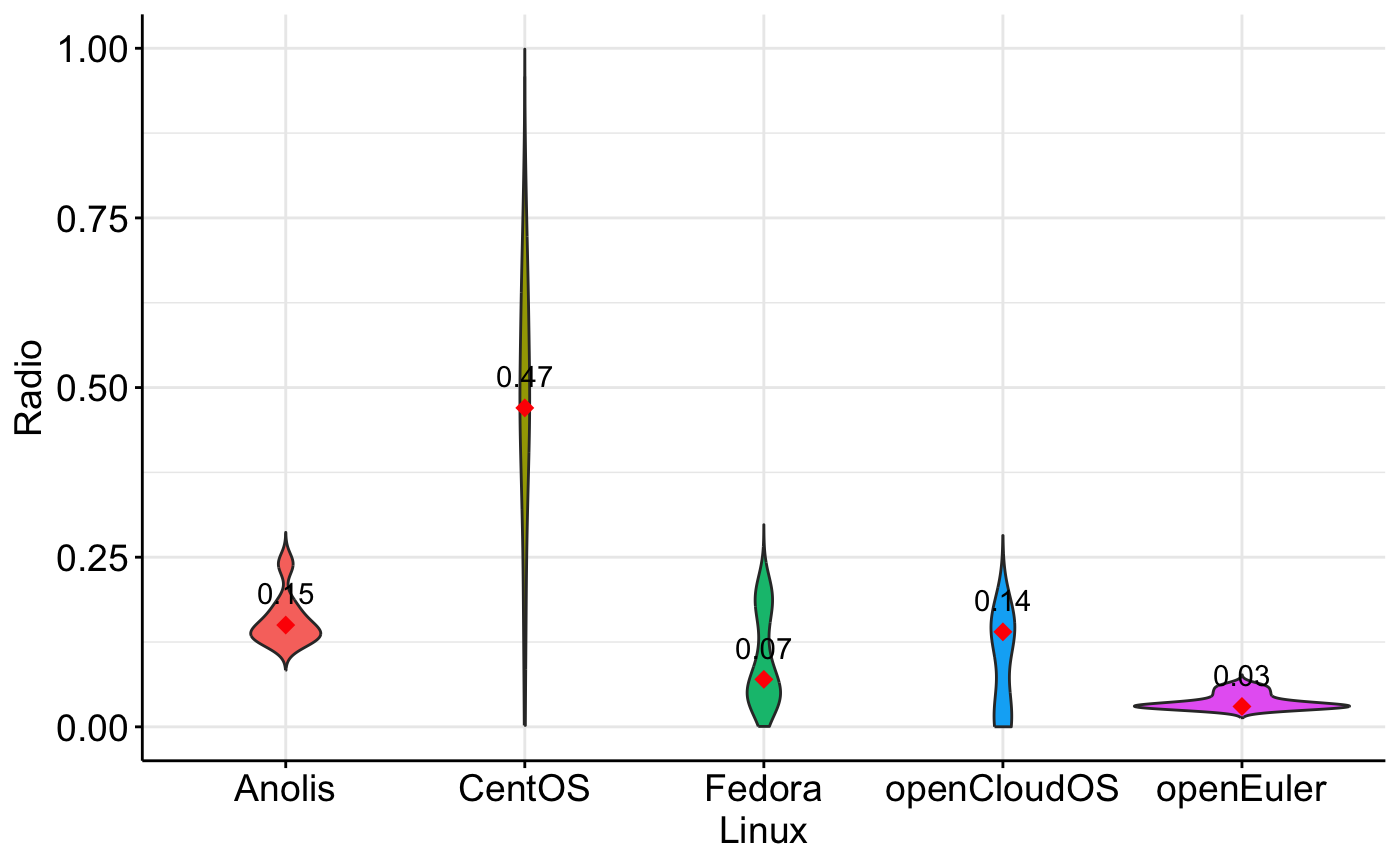}}
		\centerline{Distribution of ratio per version}
	\end{minipage}
	\vspace{-0.2cm}
	\caption{Distribution of packages applied the P2G mechanism}
	\label{rq1-distribution}
\end{figure}

\textbf{Distribution of packages adopting P2G.} Figure \ref{rq1-distribution} shows violin plots with the distribution and ratio of packages adopting P2G. \ding{182} The number of packages adopting P2G varies significantly across distributions. Fedora has a median of 3,544 packages adopting P2G and OpenEuler has only 776, which is approximately one-fifth of Fedora's. \ding{183} The distribution of package numbers across versions also varies among Linux distributions. Anolis shows significant fluctuations, ranging from 0 to 4,000, while OpenEuler remains within a more stable range of 700 to 800, indicating minimal volatility. \ding{184} The number and ratio of packages adopting P2G are not consistent. Fedora has the highest number of grouped packages (median of 3,544), but a lower ratio of 0.27. In contrast, CentOS has fewer grouped packages (a median of 1,297), but it has the highest ratio at 0.47.

\begin{figure}
    \centering
    \begin{minipage}{0.85\linewidth}
		\centerline{\includegraphics[width=\textwidth]{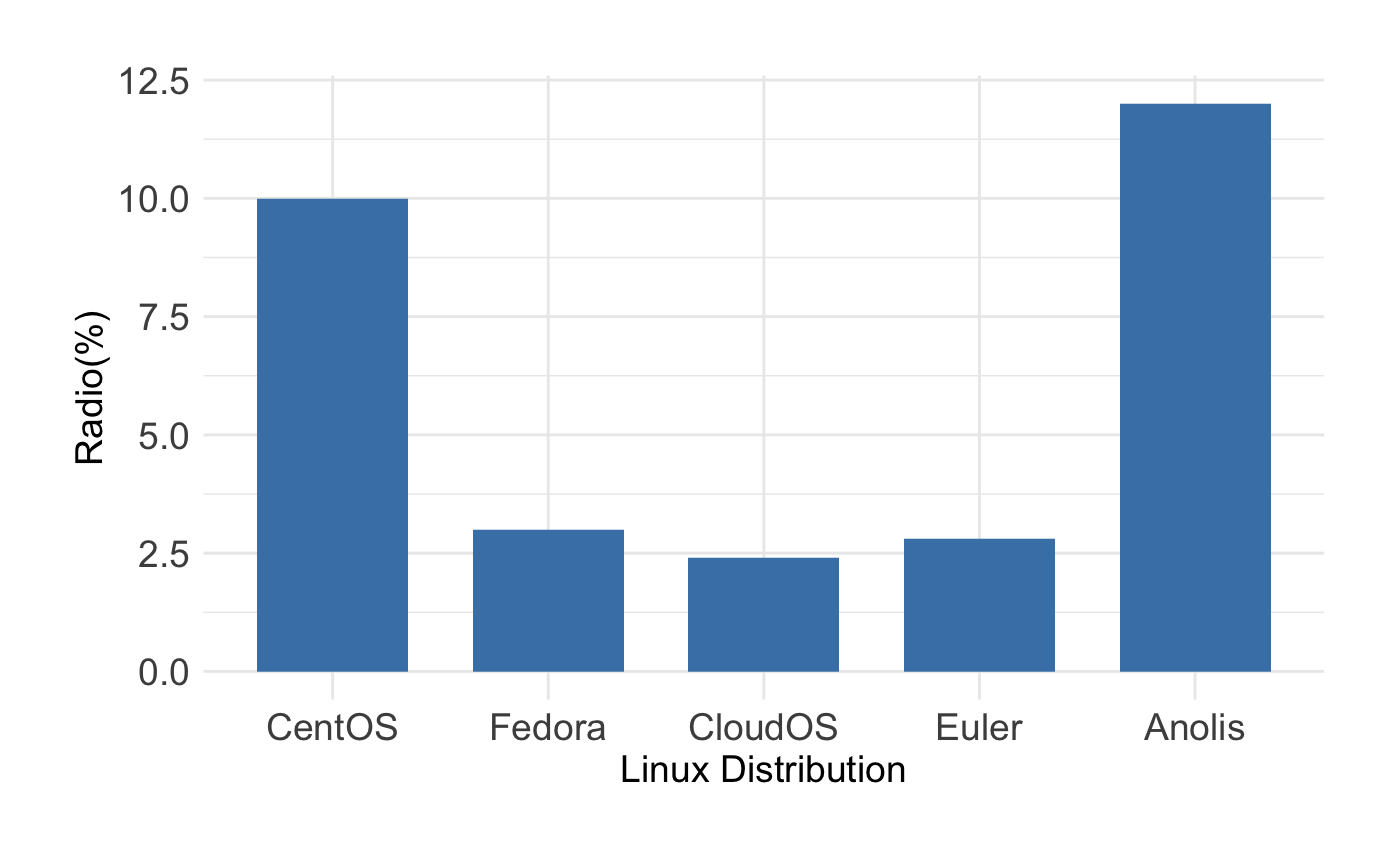}}
    \end{minipage}
    \caption{Ratio of the packages applied P2G mechanism. All the Linux distributions are sorted in descending order of popularity (\#stars).}
    \label{rq1-popularity}
\end{figure}

\textbf{Relationship between the popularity of Linux distributions and the ratio of packages adopting P2G.} As shown in Figure \ref{rq1-popularity}, the ratio shows a downward trend, except for Anolis. It suggests that a Linux distribution's popularity is correlated with the application of P2G. To validate this assumption, we perform a Spearman correlation analysis between the ratio of packages adopting P2G and the number of stars a Linux distribution has. However, the results (p-value=0.6, rho=0.4) indicate a weak positive correlation and the high p-value suggests that this correlation is statistically insignificant. Thus, it shows that the factors related to a Linux distribution's popularity are complex. The anomaly observed in Anolis may be due to its inheritance of the CentOS ecosystem, resulting in a higher ratio despite its relatively low popularity.

\begin{resultRQ}{Summary for RQ1:}
Linux distributions continuously add new packages to provide enhanced functionality and more packages are adopting the P2G mechanism, especially for popular Linux distributions. However, the packages that adopt this mechanism still account for a small proportion of the distributions' total packages.
\end{resultRQ}
\section{RQ2:Evolution Patterns} \label{sec:rq2}
We aim to explore how the P2G mechanism evolves across different versions of Linux distributions.
\subsection{Methodology}
Targeting the above goal, our method consists of two parts: a content-based summary of group change patterns and a quantitative analysis of package flow directions.

\begin{figure*}
    \centering
    \begin{minipage}{0.78\linewidth}
		\centerline{\includegraphics[width=\textwidth]{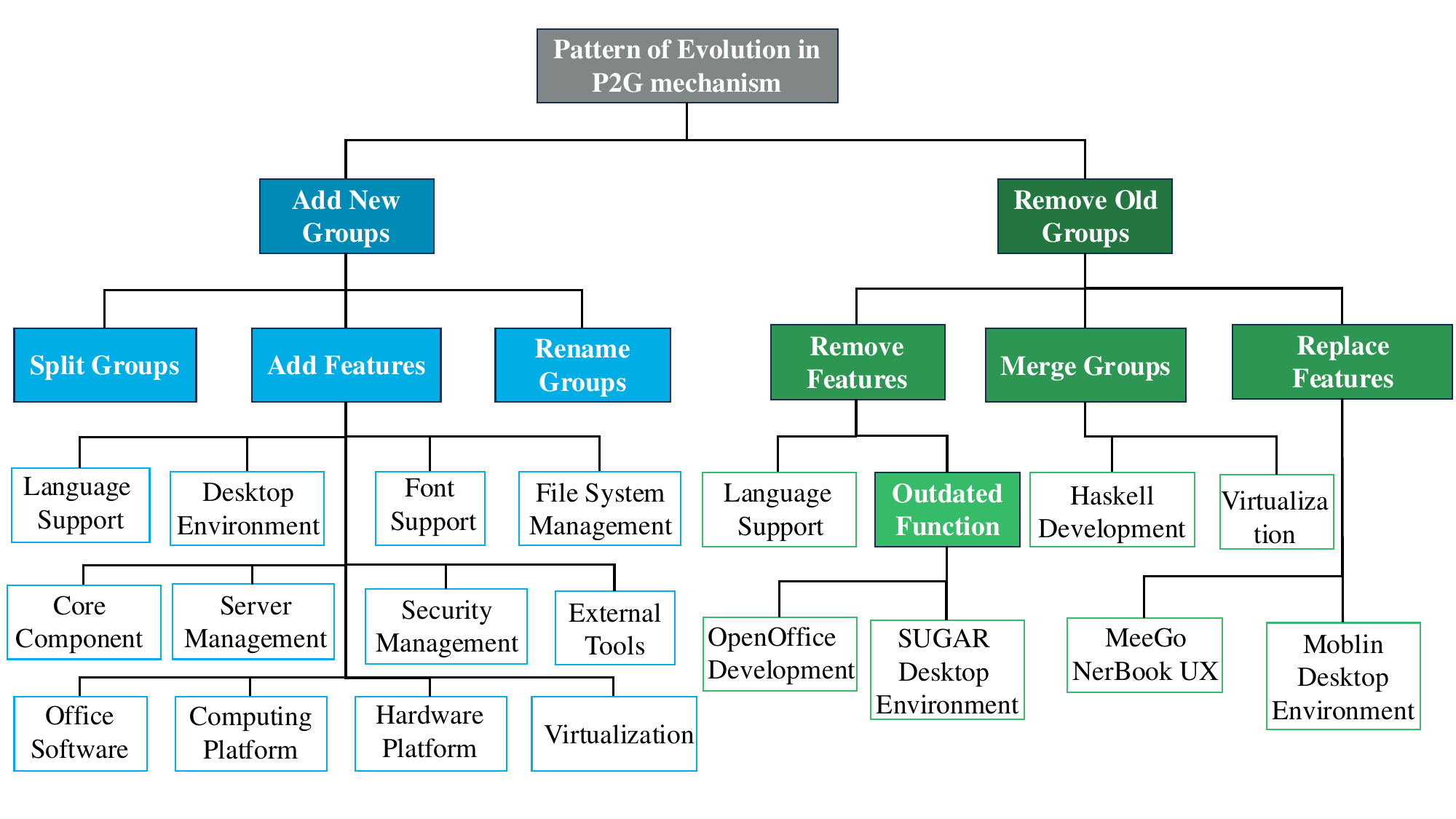}}
    \end{minipage}
    \caption{The Pattern of Evolution in P2G mechanism}
    \label{fig:rq2-pattern}
\end{figure*}

\subsubsection{\textbf{Change Patterns}} We manually review the open-source repositories of the Linux distributions mentioned in Section \ref{sec:linux} to locate the URLs of the source files containing group and package information. Then we use the \textit{request} library to scrape group information for each release version. Finally, we utilize the group information to calculate the added and removed groups between consecutive versions, including their names, functional descriptions, and contained packages.

To summarize the change patterns, we employ a manual annotation process conducted by a team. The team consists of one PhD candidate and two master students majoring in computer science. All of them are fluent in English and have completed a semester course on operating systems. To guarantee the quality of the annotation process, we set a two-round labeling process. In the first round, annotators independently review the added and removed groups, labeling each change with a descriptive pattern. Based on these individual annotations, the team can develop a taxonomy of change patterns. In the second round, the taxonomy is relabeled by two master students. In cases of disagreement, the PhD candidate is consulted to reach an agreement. 

\subsubsection{\textbf{Flow Analysis}} Packages in Linux distributions can be divided into two categories: those using P2G and those that do not. Thus, the flow sources consist of two types: packages that previously did not use P2G but now do (\ie \textit{S1}) and new packages added to the distributions (\ie \textit{S2}). The flow outcomes also fall into two categories: packages that remain in Linux but no longer use P2G (\ie \textit{O1}) and packages that are completely removed from distributions (\ie \textit{O2}).

To calculate the number of packages in each scenario, we apply the following algorithm. For each distribution version, we identify the added and removed packages (\ie $ap, rp$) under P2G compared to its previous version, as well as the total number of packages in the previous and current versions. For each package in $ap$, we check whether it exists in the previous version. if it does, it belongs to \textit{S1}, and if not, it falls under \textit{S2}. For each package in $rp$, we verify its presence in the current version. If it is found, it belongs to \textit{O1}, otherwise, it is classified as \textit{O2}. Using this algorithm, we can calculate the number of packages in four scenarios for each version. Finally, we merge them to obtain the package flow results.

\subsection{Results}

\subsubsection{\textbf{Change Patterns}} Figure \ref{fig:rq2-pattern} presents the taxonomy of change patterns in Linux distributions. There are three primary patterns for adding groups. \textbf{\ding{182} Split Groups}: This typically occurs when a group is too large or its functions too concentrated, making management inconvenient. For instance, the \textit{Chinese Support} group is split into two distinct groups(\ie \textit{Simplified Chinese} and \textit{Traditional Chinese}) from Fedora 14 to Fedora 15. \textbf{\ding{183} Add Features}: As technology advances and new requirements arise, additional groups are created to accommodate these changes. This pattern includes 12 subtypes, \eg language support, font support, and file system management. For example, Fedora increases support for different languages and regions to enhance internationalization from version 4 to 5. \textbf{\ding{184} Rename Groups}: This generally reflects technological changes or a shift in marketing strategy to ensure the group name aligns with its new function. For example, Fedora renamed \textit{OpenOffice.org Development} group to \textit{LibreOffice Development}. In addition, there are three patterns for removing groups. \textbf{\ding{185} Remove Features}: This can be categorized into two subtypes. The first occurs when a specific technology becomes outdated or is replaced by more efficient solutions, such as removing \textit{MeeGo NerBook UX Environment}. The second subtype involves the elimination of features that are no longer necessary, such as support for a specific language or font. \textbf{\ding{186} Merge Groups}: To optimize resource management, groups with related or overlapping functionalities can be merged. For example, the \textit{Virtualization Client} group and \textit{Virtualization Hypervisor} group are combined into a unified \textit{Virtualization} group. \textbf{\ding{187} Replace Features}: a group may be replaced to cover a broader or more updated technological scope. For instance, the \textit{Moblin Desktop Environment} group is integrated into the more comprehensive \textit{Virtualization} group.

To ensure the taxonomy is comprehensive and representative, we conducted a validation survey involving 12 Linux practitioners and 2 researchers. We list each pattern with its description and two examples, inviting participants to answer whether they agree with it. The results are summarized in Table \ref{tab:rq2-valid}. For each pattern, we report the percentage of ``yes'' and ``no'' answers. The most approved pattern is merging existing groups, with 100\% of ``yes'' answers. The least approved category is rename groups, which has been agreed by 40\% of these participants (a non-negligible fraction of all the participants). Overall, the average rate of ``yes'' answers across all patterns is 70\%, indicating that the patterns align with the experience of most participants (only one pattern is below 50\%).

\begin{table}[]
    \centering
    \caption{Validation Survey Results}
    \label{tab:rq2-valid}
    \begin{tabular}{cccc}
\hline
\multirow{2}{*}{\textbf{Type}}                                                         & \multirow{2}{*}{\textbf{Pattern}} & \multicolumn{2}{c}{\textbf{Response}} \\ \cline{3-4} 
                                                                              &                          & Yes           & No           \\ \hline
\multirow{3}{*}{\begin{tabular}[c]{@{}c@{}}Add  Groups\end{tabular}}   & Add Features         & 80\%          & 20\%         \\
                                                                              & Split Groups    & 80\%          & 20\%         \\
                                                                              & Rename Groups            & 40\%          & 60\%         \\ \hline
\multirow{3}{*}{\begin{tabular}[c]{@{}c@{}}Remove Groups\end{tabular}} & Remove Features          & 60\%          & 40\%         \\
                                                                              & Merge Groups    & 100\%         & 0\%          \\
                                                                              & Replace Features & 60\%          & 40\%         \\ \hline
\end{tabular}
\end{table}

\begin{figure}
    \centering
    \begin{minipage}{0.85\linewidth}
		\centerline{\includegraphics[width=\textwidth]{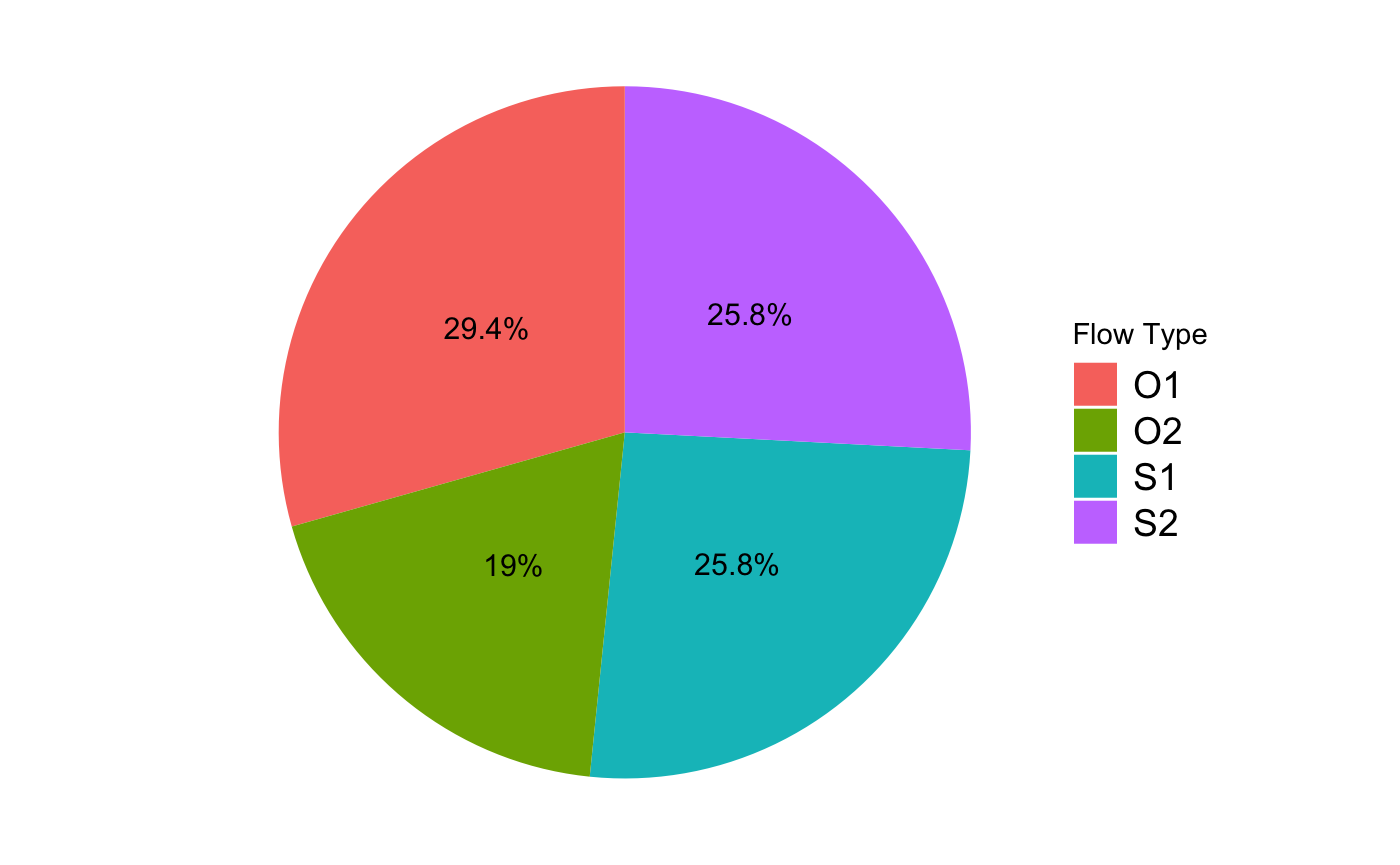}}
    \end{minipage}
    \caption{Flow of packages in P2G mechanism}
    \label{fig:flow}
\end{figure}

\subsubsection{\textbf{Flow Analysis}}  Figure \ref{fig:flow} shows the percentage of each flow situation. \ding{182} We can see that the number of packages added to groups surpasses those removed. Added packages from both flow sources make up 51.6\%. This suggests that with each version update, more packages are starting to adopt P2G. \ding{183} Two types of flow sources for P2G have the same percentage. Half of the packages come from the Linux distributions themselves (\ie \textit{S1}) and the other half comes from external sources (\ie \textit{S2}). \ding{184} Packages that no longer use P2G are more likely to remain in distributions. Among these, 29.4\% have stayed in distributions (\ie \textit{O1}) while 19\% have been directly removed from them (\ie \textit{O2}).

\begin{resultRQ}{Summary for RQ2:}
The P2G in Linux distributions has six major change patterns, including splitting and merging groups, etc. With version evolution, more packages adopt P2G and those no longer using P2G often remain in distributions. 
\end{resultRQ}

\section{RQ3:Quality Assessment} \label{sec:rq3}
We aim to evaluate the quality of groups under the P2G mechanism in the latest distributions and identify potential issues.

\subsection{Methodology}

\subsubsection{\textbf{Proposed Metrics}} Based on feedback from P2G practitioners, an ideal group should meet the following criteria:
\begin{itemize}
    \item \textbf{Compactness}: Since packages within a group should work together to support a common feature, they must be functionally related or interdependent.
    \item \textbf{Relevance}: The description of a group is critical in guiding developers to choose packages, so it should summarize the functionality of the packages accurately.
    \item \textbf{Differentiation}: Groups should have a distinct purpose, and low differentiation between them could confuse developers when selecting groups. 
    \item \textbf{Distributions}: the number of packages in a group should be balanced. Overcrowded groups can lead to confusion when retrieving specific items, while groups with too few packages may result in inefficient management.
\end{itemize}

\begin{figure}
    \centering
    \begin{minipage}{\linewidth}
		\centerline{\includegraphics[width=\textwidth]{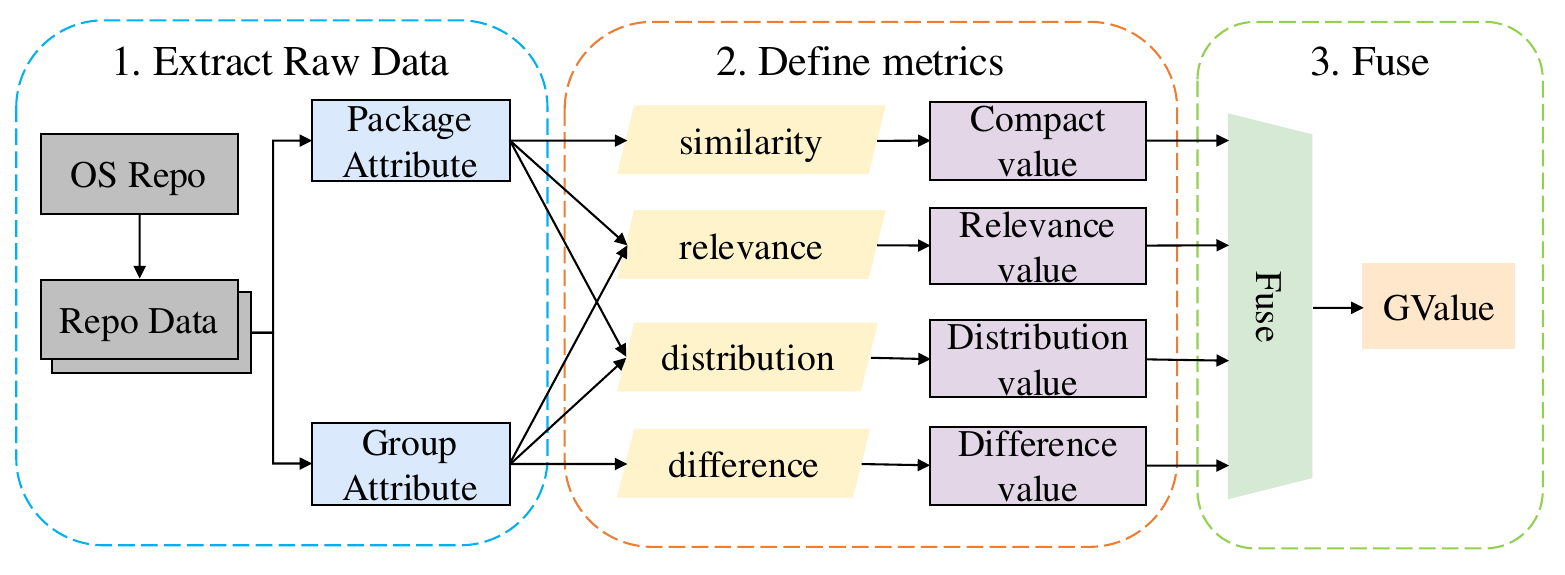}}
    \end{minipage}
    \caption{Overview of \textit{GValue}}
    \label{fig:gvalue}
\end{figure}

Based on these expert criteria, we propose \textit{GValue}, a metric for evaluating the quality of groups in P2G. Figure \ref{fig:gvalue} provides an overview of our \textit{GValue}, which consists of three steps. In the first step, two types of information are extracted from the distribution repositories, which is package attributes and group attributes. In the second step, they are processed and transformed into a set of values that reflect the quality of groups according to the four expert criteria. Finally, these values are consolidated into a single metric. The following sections provide a detailed description for each step.

\textbf{Extract raw data.} As illustrated in figure \ref{fig:gvalue}, the step aims to extract group and package attributes from the distribution repositories. We first scrape $n$ groups from repositories, denoted as $O = \{g_1,g_2,...,g_n\}$. Then, three attributes are extracted for each group $g_i$, including name, description, and package list, expressed as $g_i = \{gn_i,gd_i,pl_i\}$. Assuming group $g_i$ contains $m$ packages, the package list is represented as $pl_i = \{p_{i1},p_{i2},...,p_{im}\}$. Additionally, we extract package attributes for each package in groups, including package name and package description, which can be defined as $p_{ij} = \{pn_{ij},pd_{ij}\}$. 

\textbf{Compactness Value.} \textit{Compact} refers to the functional similarity or dependency degree among packages within a group. To measure similarity, we leverage the rich semantic information contained in package function descriptions. Specifically, we utilize a pre-trained language model ($PLM$) to convert these descriptions into semantic vectors and calculate cosine similarity between them as similarity scores. 

\begin{equation} \label{sim} 
\begin{aligned}
    V_{ij} = PLM(d_{ij}), V_{ik} = PLM(d_{ik}) \\
    sim(p_{ij},p_{ik}) = \frac{V_{ij} \cdot V_{ik}}{\|V_{ij}\| \|V_{ik}\|}
\end{aligned}
\end{equation}
where $V_{ij}$ represents the semantic vectors of the description for the $j$-th package in the $i$-th group and $Sim(p_{ij},p_{ik})$ denotes the functional similarity between the $j$-th package and $k$-th package in the $i$-th group.

Then, we quantify the dependency degree by the path length in the package dependency graph. It can reflect the relationships between packages effectively, as a shorter path in the graph indicates a higher degree of dependency.

\begin{equation} \label{dep}
    dep(p_{ij},p_{ik}) = \frac{1}{d(p_{ij},p_{ik})} 
\end{equation}
where $d(p_{ij},p_{ik})$ and $dep(p_{ij},p_{ik})$ denotes the path length and dependency degree between the j-th packages and the k-th package in the i-th group, respectively.

Finally, since functional similarity and dependency degree are two relatively independent aspects, we take the larger value between them. Thus, we first calculate the similarity score (\ie equation \ref{sim}) and dependency degree (\ie equation \ref{dep}) between any two packages in a group and then take the average as the compact value for a group.

\begin{equation}
\begin{aligned}
    com(g_i) &= \frac{\sum_{j=1}^{m} \sum_{k=i}^{m} max(sim(p_{ij},p_{ik}),dep(p_{ij},p_{ik}))}{m(m-1)}
\end{aligned}
\end{equation}
where $com(g_i)$ denotes the compact value for the i-th group.

\textbf{Relevance Value.} 
To calculate the relevance between the descriptions of a group and packages in the group, we first input the group description and package descriptions into $PLM$ to obtain their semantic representation (\ie $V_i = PLM(gd_i)$). Then, we compute the cosine similarity between them (\ie equation \ref{sim}). Last, we take the average of these similarity scores as the overall relevance value for a group.

\begin{equation}
    \begin{aligned}
        rel(g_i) &= \frac{\sum_{j=1}^m sim(gd_i, p_{ij})}{m}
    \end{aligned}
\end{equation}
where $rel(g_i)$ represents the relevance value for the i-th group.

\textbf{Differentiation Value.} 
A group has three main attributes: name, description, and a list of packages. To comprehensively assess the differences between groups, we evaluate each attribute individually before combining them into an overall score. Specifically, we design methods to quantify the differences for each attribute individually, which are then integrated to provide a comprehensive assessment.

Firstly, edit distance is chosen to measure the difference between names. This choice stems from our observation that some group names are simply character strings, making it difficult to extract semantic information, such as the group \textit{KDE} in CentOS 7.
\begin{equation}
    \begin{aligned}
        ndif(g_i,g_j) &= \frac{ED(gn_i,gn_j)}{max(\|gn_i\|,\|gn_j\|)} \\
        ndif(g_i) &= \frac{\sum_{\substack{j=1 \\ j \neq i}}^n ndif(g_i, g_j)}{n - 1}
    \end{aligned}
\end{equation}
where $ED(gn_i,gn_j)$ and $ndif(g_i,g_j)$ represent the edit distance and difference score between the i-th group's name and the j-th group's name, respectively. $\|gn_i\|$ denotes the string length of the i-th group's name and $ndif(g_i)$ refers to the different score between the name of i-th group and all other groups.

To assess the differences between group descriptions, we also use $PLM$ to compute the semantic vector ($V_i$) for each group and measure the difference using cosine similarity. Since cosine similarity ranges from -1 to 1 and larger values indicate smaller differences, we apply interval shifting to normalize the values, restricting them to a range between 0 and 1~\cite{may2009development}. 

\begin{equation}
    \begin{aligned}
        sim(g_i,g_j) &= \frac{V_i \cdot V_j}{\|V_i\| \|V_j\|} \\
        ddif(g_i,g_j) &= 1 - \frac{(sim(g_i,g_j) +1)}{2} \\
        ddif(g_i) &= \frac{\sum_{\substack{j=1 \\ j \neq i}}^n ddif(g_i, g_j)}{n - 1}
    \end{aligned}
\end{equation}
where $sim(g_i,g_j)$ and $ddif(g_i,g_j)$ refer to the similarity score and difference score between the descriptions of the i-th group and the j-th group, respectively. $ddif(g_i)$ represents the average difference score between the descriptions of i-th group and all other groups.

Lastly, since the package list in a group can be viewed as a set, set similarity can be used as a metric to measure the difference between package lists. However, considering that the packages have varying levels of necessity(\eg \textit{mandatory}, \textit{default}, and \textit{optional}), treating all packages as equally important and applying set similarity directly is not appropriate. To address this, different weights are assigned to each type of package. The principle behind the weight assignment is that the more essential a package is, the higher its weight. $w_i(p_j)$ is the weight function for the packages. The determination of this function is based on multiple iterations through computing its correlation with manual scoring.
\begin{equation}
    w_i(p_j) = 
\begin{cases} 
0.8 & \text{if } p_j \in mandatory \\ 
0.5 & \text{if } p_j \in default \\ 
0.2 & \text{if } p_j \in optional 
\end{cases}
\end{equation}

Based on the weight function and Jaccard similarity~\cite{ivchenko1998jaccard}, we propose a quantitative method to measure the differences among package lists in distinct groups.
\begin{equation}
\begin{aligned}
    pdif(pl_i,pl_j) &= 1- \frac{\sum_{x \in pl_i \cap pl_j} min(w_i(x),w_j(x))}{\sum_{x \in pl_i \cup pl_j} max(w_i(x),w_j(x))} \\
    pdif(g_i) &= \frac{\sum_{\substack{j=1 \\ j \neq i}}^n pdif(pl_i, pl_j)}{n - 1}
\end{aligned}
\end{equation}
where $pdif(pl_i,pl_j)$ represents the difference score between the package list in the i-th and j-th group. $pdiff(g_i)$ denotes the difference between the package list in the i-th group and all other groups.

After measuring the differences for each attribute, we combine them and calculate the average to obtain the overall score (\ie $dif(g_i)$), representing a group difference score.

\textbf{Distribution Value.} The distribution value reflects the rationality of the number of packages in a group. According to expert experience, a reasonable number should fluctuate around a certain mean, with excessively high or low values being considered anomalies. Therefore, we can assume that the number of packages in each group follows a normal distribution~\cite{krithikadatta2014normal}. Based on this assumption, the mean ($\mu$) and standard deviation ($\sigma$) can be calculated, and an appropriate range (\ie $\mu \pm 2\sigma$) can be selected as the upper and lower bounds for a reasonable range.  If the number of packages falls within the reasonable range (\ie $[\mu - 2\sigma, \mu + 2\sigma]$), we assign the distribution value as 1. Otherwise, we assign it as 0.

\textbf{GValue Fuse.} Based on the four dimensions score, we calculate the average value (\ie $GValue(g_i)$) as the comprehensive evaluation metric for assessing the quality of a group. 

\subsection{Results}

\subsubsection{\textbf{Metrics Validation}}
\begin{figure}
    \centering
    \begin{minipage}{0.99\linewidth}
		\centerline{\includegraphics[width=\textwidth]{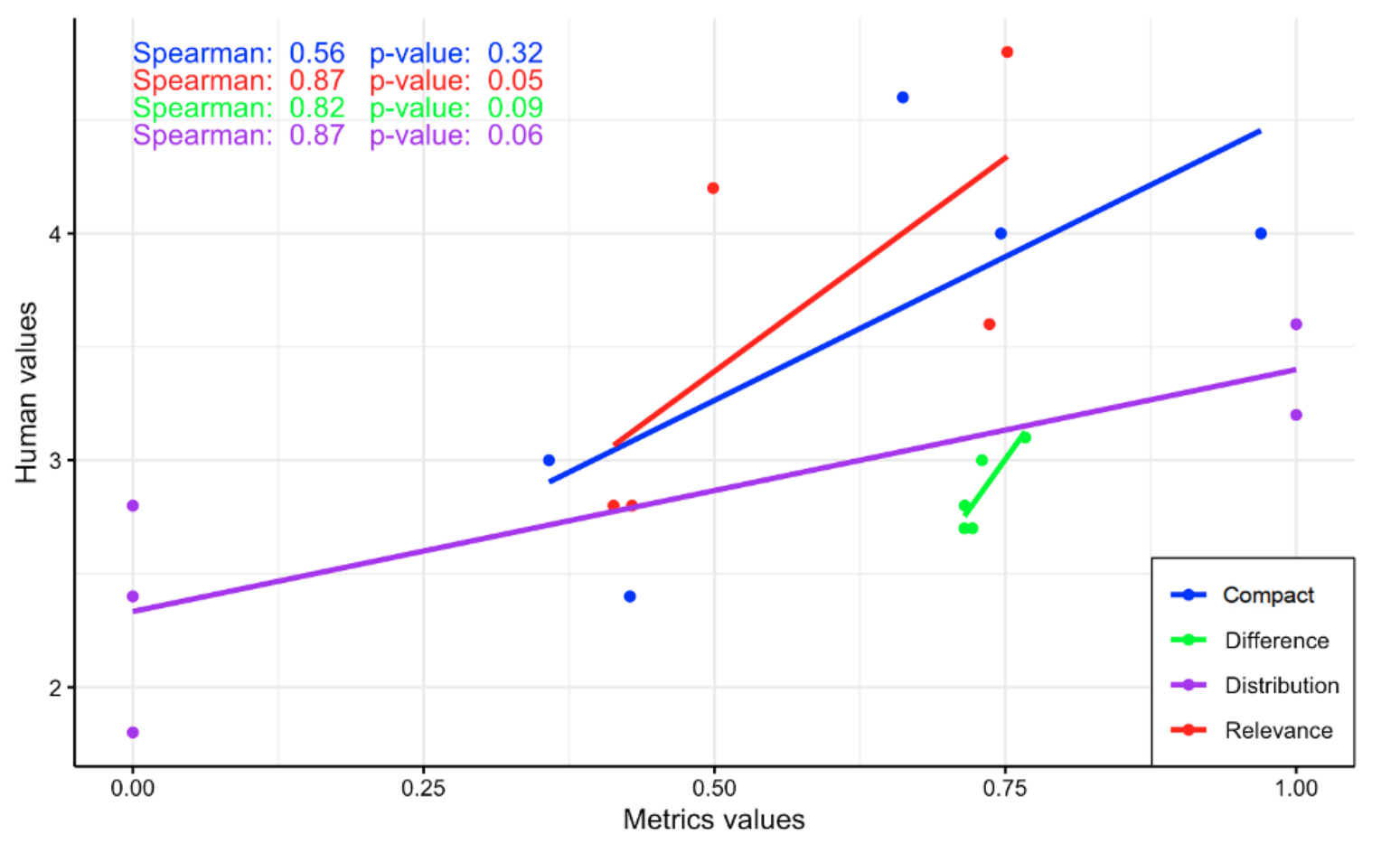}}
    \end{minipage}
    \caption{Correlation of {\sc GValue} and human score}
    \label{fig:rq3-validate}
\end{figure}

We validated our {\sc GValue} metric by evaluating five groups randomly selected from five different Linux distributions. The ground truth for these groups was established through a survey completed by 16 skilled participants. Then we calculate the Spearman correlation~\cite{de2016comparing} between the human score and {\sc GValue} score as an evaluation metric. The result is shown in Figure \ref{fig:rq3-validate} and indicates the {\sc GValue} have a strong correlation with human score ($rs > 0.7$) in each aspect.

\subsubsection{\textbf{Problems Analysis}}  
Using the \textit{GValue} metric, we evaluated the quality of groups in the latest versions of five Linux distributions and found that 16\% of the groups scored below 0.2, indicating significant room for improvement. After manually reviewing these low-quality groups, we identified various issues in adopting this mechanism, involving group-related problems (e.g., inappropriate descriptions) and package-level issues (e.g., insufficient numbers). Full details can be found in our open-source dataset~\cite{web:code}.

\begin{resultRQ}{Summary for RQ3:}
We develope a multi-dimensional feature-based method to evaluate group quality, which shows a strong correlation with human assessments. Besides, our analysis reveals that 16\% of the groups exhibit poor quality, with two types of issues in the adoption of P2G.
\end{resultRQ}

\section{RQ4:Group Tendency} \label{sec:rq4}
We aim to identify which types of packages are more inclined to adopt the P2G mechanism.

\subsection{Methodology}
We first conduct a topic analysis on group descriptions to understand the tendency from functionality. Then, we compare the key information between the descriptions of packages in groups and those not assigned to any groups.

\subsubsection{Topic Analysis on group descriptions} Latent Dirichlet Allocation~\cite{blei2003latent} (LDA) is a widely used unsupervised learning method for extracting topics from large-scale unstructured text. By mining the latent semantic structure of the text, LDA enables topic modeling across a collection of descriptions to identify common themes. Thus, we use LDA to conduct topic analysis on the group descriptions. Specifically, we start with preprocessing for the group descriptions, including tokenizing and removing stopwords using NLTK~\cite{hardeniya2016natural}. We then construct a vocabulary dictionary using the Gensim~\cite{vrehuuvrek2011gensim} and generate a group-word frequency matrix, representing the frequency of each word in each group. Finally, each topic is characterized by the top ten words with the highest weights, revealing the latent semantic structure in the group descriptions. Additionally, as LDA requires specifying the number of topics in advance, we perform a series of experiments with various numbers of topics and evaluate them using the coherence score to determine the optimal number. 

\subsubsection{Extraction of Key Information from package descriptions} 
We extract and compare the key information from the descriptions of the packages in groups with those that are not. Package descriptions often contain many redundant and trivial words, so we perform tokenization and stopword removal. Then, we use the term frequency-inverse document frequency (TF-IDF) algorithm~\cite{ramos2003using} to mine keywords. The TF-IDF algorithm determines the importance of a word by comparing its frequency in a specific package description against its rarity across all package descriptions. The algorithm can measure how relevant a word is within a given package description. Specifically, for a set of package descriptions $S$, a specific package description $p$ from this set, and a word $w$ in $p$, the relevance $i_w$ of the word is calculated as follows:

\begin{equation}
    i_r = f_{w,p} \log\left(\frac{|S|}{f_{w,S}}\right),
\end{equation}
where $f_{w,p}$ represents the frequency of $w$ in $p$, $\|S\|$ is the number of package descriptions in the set $S$, and $f_{w,S}$ denotes the frequency of $w$ across the entire set $S$. Using this algorithm, we extract the most relevant words from each package description. To help developers determine whether to apply P2G to a package, we collect and compare the keywords from package descriptions within and outside of groups. 

\subsection{Results}

\begin{figure}
    \centering
    \begin{minipage}{0.9\linewidth}
		\centerline{\includegraphics[width=\textwidth]{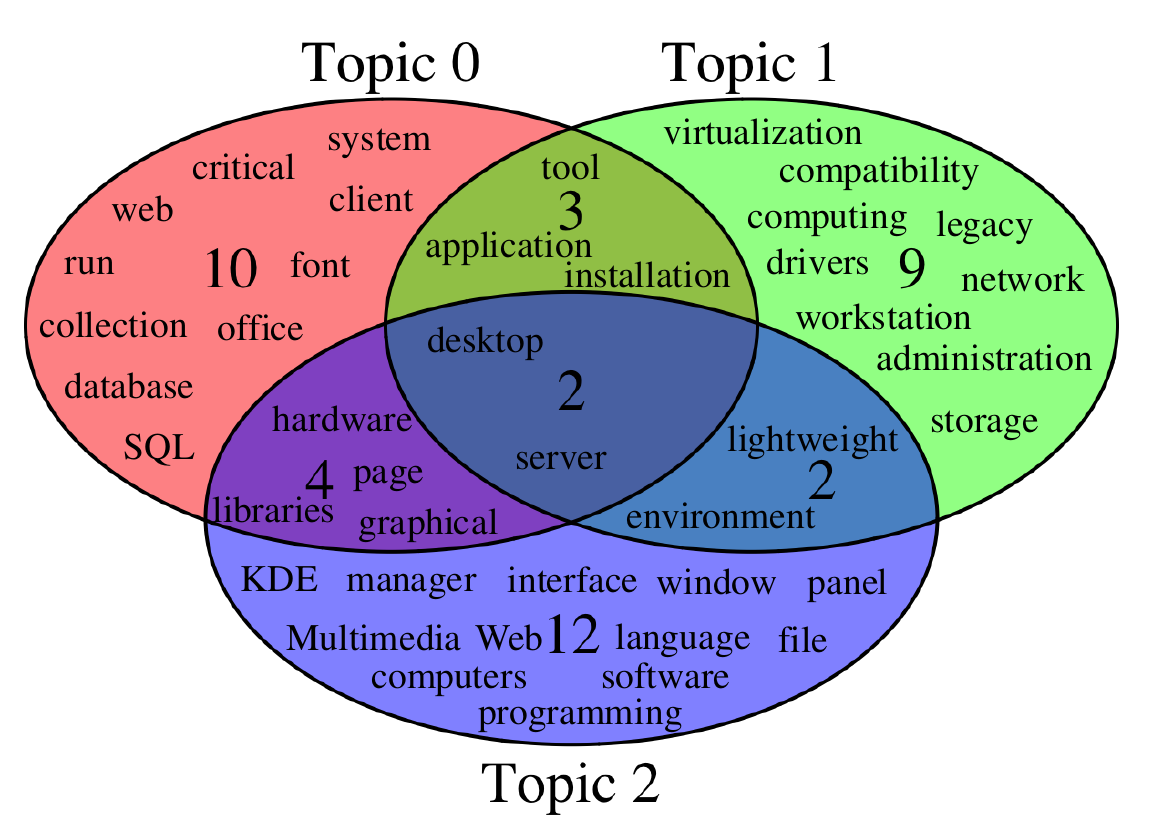}}
    \end{minipage}
    \caption{Venn Diagram of the Topic Modelling}
    \label{fig:topic}
\end{figure}

\subsubsection{Topic in group descriptions} we set the number of topics ranging from 1 to 20 and calculate their coherence scores. We find the coherence score is the highest (\ie 0.48) when the number is 3. Figure \ref{fig:topic} presents the topics for group descriptions under this situation. We can observe that they mainly involve the following categories: Applications (\eg fonts and tools), Development (\eg programming and libraries), Graphical Desktop (\eg desktop and KDE), Networks (\eg network and virtualization), and Servers (\eg servers). This suggests that developers tend to adopt P2G when a package's functionality belongs to one of these categories.

\subsubsection{Key information in package descriptions} 
\begin{figure}
    \centering
    \begin{minipage}{0.9\linewidth}
		\centerline{\includegraphics[width=\textwidth]{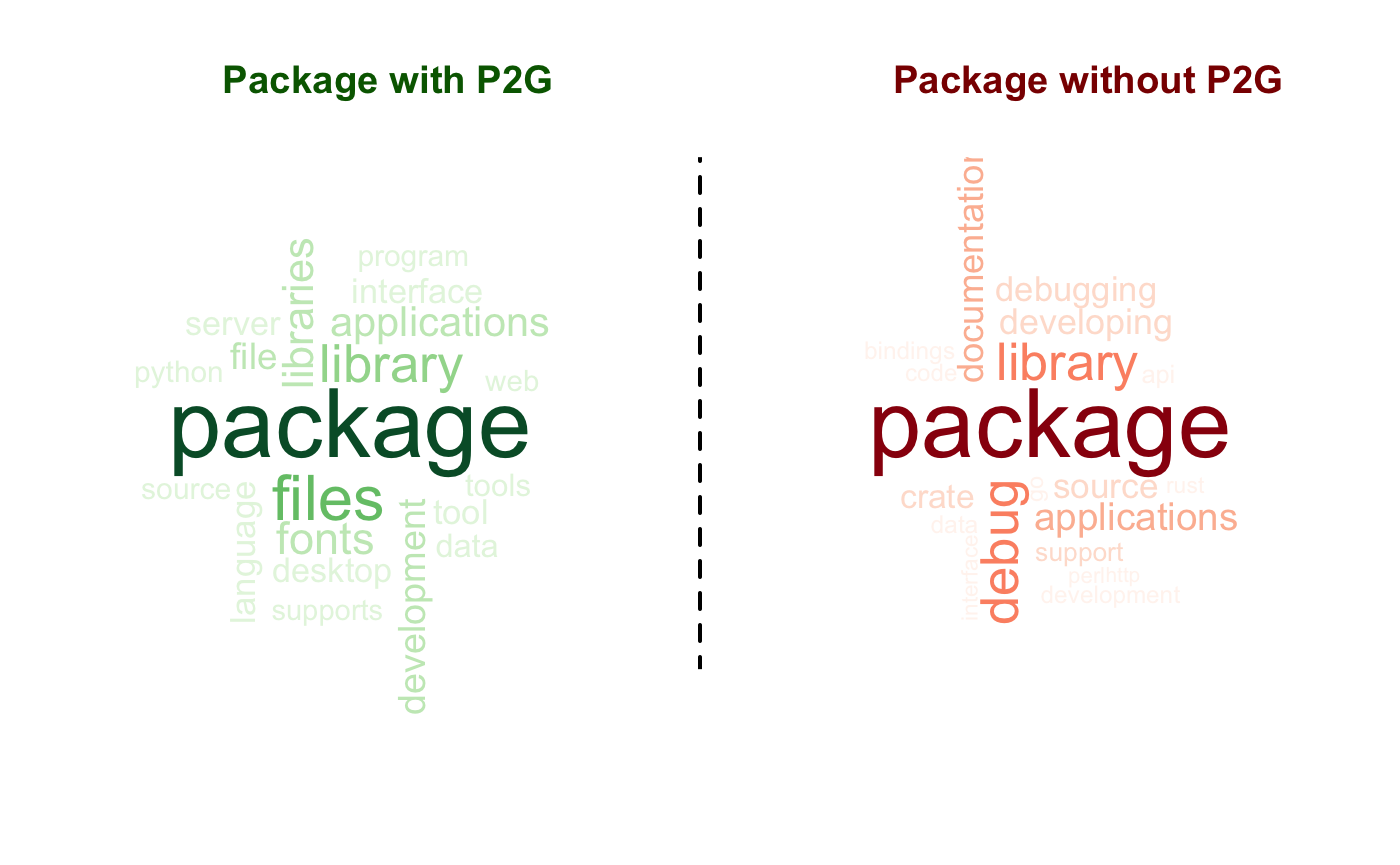}}
    \end{minipage}
    \caption{Comparison of the keywords in packages with P2G and those without.}
    \label{fig:compare}
\end{figure}

Figure \ref{fig:compare} shows the keywords that can distinguish packages with P2G from those without it, presented in two clouds. We can observe that the keywords of packages with P2G emphasize general utility and development resources, \eg ``files'', ``fonts'', and ``applications''. It suggest that packages with P2G focus more on infrastructure-related tasks and provide various development environments. The packages without P2G highlight more technical aspects, with keywords like ``debug'' and ``documentation''. This suggests they focus more on debugging tools and development documentation.

\begin{resultRQ}{Summary for RQ4:}
The following types of packages may be more inclined to adopt the P2G by Linux developers: 1) General Applications (\eg fonts and languages), 2) Infrastructure Support (\eg graphical desktop and network), and 3) Development Resources (\eg programming languages and tools).
\end{resultRQ} 
\section{Discussion}

\subsection{Implications}
We offer practical implications of our findings in this comprehensive empirical study.

\textbf{Guidance for Developers and Maintainers}: By identifying the categories of packages that commonly adopt P2G (\eg graphical desktop and network), developers and maintainers can better predict and decide which packages should adopt P2G in future versions. This categorization can provide practical guidelines for them.

\textbf{Support for OSS Project Evolution}: Summarizing the evolution patterns of the P2G mechanism, such as splitting and merging groups, enables developers to predict potential changes and manage updates more effectively. The conclusion that packages that no longer adopt P2G are likely to remain within distributions underscores the resilience and long-term stability of OSS projects.

\textbf{Enhanced Quality Control}: The {\sc GVALUE} provides a concrete method to assess the quality of groups within OSS projects. Identified issues (\eg inappropriate descriptions and insufficient numbers) will help developers and maintainers to improve the quality of groups. The {\sc GVALE} and insights can contribute to control and optimize groups and improve long-term project success. 

\subsection{Threat to validity}
\textbf{Construct Validity} concerns the relationship between the treatment and the outcome. The threat comes from the rationality of the research questions we asked. We are interested in the package-to-groups mechanism in Linux distributions. To achieve this goal we focus on its application trend, evolution pattern, quality assessment, and group tendency. We believe that these questions have a high potential to provide unique insights and value for practitioners and researchers.

\textbf{Internal Validity} addresses potential threats to the way the study was conducted. The first threat relates to the identification of packages and groups. This was done manually, which may have resulted in some omissions, but ensured the overall quality of the dataset. The second threat concerns the manual summary of group change patterns when answering \hyperref[sec:rq2]{RQ2}. To address this, we employ an experienced team and set a two-round labeling process. The third threat relates to the topic extracted through topic modeling. We acknowledge that the choice of these topics is to some subjective. To mitigate this threat, we invite two Linux practitioners to conduct the topic selection. After this initial proposal, we compare our topics and revised the results.

\textbf{External Validity} considers the generalizability of our findings. This study primarily focuses on popular open-source Linux distributions. Although the P2G mechanism is important across all OSS projects, our research reveals that it is rarely implemented in less popular projects (Section \ref{sec:rq1}). Furthermore, developers working on these less active projects often face time constraints, which may hinder the adoption of P2G. This also explains the limited application in these cases.


\section{Conclusion}
In this paper, we conduct a preliminary study on the package-to-group mechanism from its application trend, evolution patterns, group quality, and group tendency. We find that although the mechanism has become more prevalent, especially in popular distributions, it still represents a small proportion of the total packages. We also summarize six major change patterns of P2G and find packages that no longer use P2G tend to remain in Linux distributions. Additionally, we propose the {\sc GValue} metric to evaluate the quality of groups, finding that many groups still exhibit poor quality. We also summarize types of packages that may be more inclined to adopt P2G by developers to support future updates. Our work investigates this mechanism from multiple aspects, which can provide significant insights for OSS projects developers and maintainers. 

\section{ACKNOWLEDGMENT}
This research was supported by the National Natural Science Foundation of China (62192731, 62192730, 62162051).



\useunder{\uline}{\ul}{}

\normalem
\balance
\bibliographystyle{IEEEtran}
\bibliography{main}

\end{document}